\documentclass[12pt,letter]{article}

\usepackage{graphicx, epsfig, color}
\def\eslt{ME_T}
\def\to{\rightarrow}

\def\bi{\begin{itemize}}
\def\ei{\end{itemize}}

\def\tchi{\tilde\chi}

\def\ttbar{t\bar{t}}

\def\sps1ap{SPS1a$^\prime$}
\def\c1p{C1$^\prime$}
\def\atp{$A_T^\prime$}

\def\tg{\tilde g}

\def\tell{\tilde\ell}
\def\tq{\tilde q}

\def\tz{\widetilde Z}
\def\alt{\stackrel{<}{\sim}}
\def\agt{\stackrel{>}{\sim}}
\def\be{\begin{equation}}  
\def\ee{\end{equation}}  
\def\bea{\begin{eqnarray}}  
\def\eea{\end{eqnarray}}  
\def\beas{\begin{eqnarray*}}  
\def\eeas{\end{eqnarray*}}  
\newcommand\prd[3]{{\it Phys.\ Rev.\ }{\bf D #1} (#2) #3}

\newcommand\plb[3]{{\it Phys.\ Lett.\ }{\bf B #1} (#2) #3}
\newcommand\jhep[3]{{\it J. High Energy Phys.\ }{\bf #1} (#2) #3}

\newcommand\npb[3]{{\it Nucl.\ Phys.\ }{\bf B #1} (#2) #3}
\newcommand\epjc[3]{{\it Eur.\ Phys.\ J. }{\bf C #1} (#2) #3}

\newcommand\zpc[3]{{\it Z.\ Physik }{\bf C #1} (#2) #3}
\newcommand{\hepph}[1]{hep-ph/#1}

\begin{document}
\begin{titlepage}
\begin{flushright}
FSU-HEP/080116
\end{flushright}

\vspace{0.5cm}
\begin{center}
{\Large \bf 
Early SUSY discovery at LHC without missing $E_T$:\\
the role of multi-leptons
}\\ 
\vspace{1.2cm} \renewcommand{\thefootnote}{\fnsymbol{footnote}}
{\large Howard Baer\footnote[1]{Email: baer@hep.fsu.edu },
Harrison Prosper\footnote[2]{Email: harry@hep.fsu.edu}, 
Heaya Summy\footnote[3]{Email: heaya@hep.fsu.edu}}\\ 
\vspace{1.2cm} \renewcommand{\thefootnote}{\arabic{footnote}}
{\it 
Dept. of Physics,
Florida State University, Tallahassee, FL 32306, USA \\
}

\end{center}

\vspace{0.5cm}
\begin{abstract}
\noindent 
Traditional searches for SUSY at hadron colliders rely heavily on 
the presence of large missing transverse energy ($\eslt$) 
to reject background compared to signal. 
On the other hand, initial searches for new physics at the LHC
may not be able to rely on $\eslt$ due to a variety of 
detector calibration issues.
We show that much of SUSY parameter space is accessible to
discovery even {\it without} using $\eslt$, and with rather low 
integrated luminosities 0.1-1 fb$^{-1}$. 
A key role is played by isolated lepton multiplicity which arises from
gluino and squark cascade decays. Requiring $\ge 3$ isolated leptons
plus jets yields a high rate of background rejection compared to signal. 
We find an LHC reach in $m_{\tg}$ of about 700-750 GeV 
for just 0.1 fb$^{-1}$ of integrated luminosity by requiring events with
$\ge 4$ jets plus $\ge 3$ isolated leptons but {\it without} using $\eslt$.
If a large enough event sample is assembled, then kinematic
reconstruction of sparticle mass properties should be possible just as in 
the case where large missing $E_T$ is required.
SUSY without $\eslt$ can also be seen in opposite-sign/same flavor 
{\it dilepton plus jets} events when a characteristic invariant mass edge stands out
against background.

\vspace*{0.8cm}

\end{abstract}


\end{titlepage}

It is expected that the CERN Large Hadron Collider (LHC), a 
$\sqrt{s}=14$ TeV $pp$ collider, will begin operation in late 2008 or 
early 2009. One of the main goals of the LHC is to either discover or
exclude the existence of weak scale supersymmetry (SUSY)\cite{wss}.
SUSY is an exciting concept theoretically in that the added symmetry
allows for a stabilization of the weak scale when the 
Minimal Supersymmetric Standard Model (MSSM) is embedded into
more encompassing theories such as GUTs, which would be valid at much 
higher energy scales. SUSY theories also have some experimental
support in that the $SU(3)_C\times SU(2)_L\times U(1)_Y$ gauge 
couplings measured at LEP are seen to unify at 
$M_{GUT}\simeq 2\times 10^{16}$ GeV under MSSM renormalization group
running, while unification is very poor under Standard Model (SM) running.

Most theories of weak scale SUSY have an added $R$-parity invariance
which is necessary to stabilize the proton against rapid decay
through $R$-violating interactions. A consequence of $R$-parity
conservation is that superpartners of SM particles must decay to other
superpartners. In this case, the lightest SUSY particle (LSP) must be 
absolutely stable. If produced in the early universe, then there 
should exist relic LSPs in the universe today, and in fact it is popular
to conjecture that these might make up the required cold dark matter (CDM)
in the universe. Null searches for massive charged or colored relics from the
Big Bang indicate that the LSP must be electrically and color neutral.
In many models, the lightest neutralino ($\tchi_1^0$ or $\tz_1$)
turns out to be the LSP, and is an excellent candidate CDM particle.
A neutralino LSP, if produced in a collider experiment, would
escape detection and thus provide a signal characterized 
by an apparent non-conservation of (transverse) energy.

It was recognized early on that perhaps the {\it classic} signature
for production of SUSY particles in collider events is 
the presence of an excess of $\eslt +$ jets events above SM 
background\footnote{Indeed, it is suggested in Ref. \cite{zen}
that SUSY gives rise to so-called ``zen'' events: jets balanced by
$\eslt$, which correspond to the sound of one hand clapping.}.
Thus, most studies of sparticle discovery at collider experiments
rely on the presence of large $\eslt$ in the events in order to 
reject SM backgrounds such as multi-jet production in QCD. At LHC,
many analyses require for instance $\eslt \agt 100$ GeV as a minimum
requirement\cite{SUSYatLHC}.

From the experimental side, the requirement of large $\eslt$ can be
problematic, especially if an {\it early} discovery of SUSY is desired.
Missing transverse energy can arise not only from the presence of 
weakly interacting neutral particles such as neutrinos or the lightest
neutralinos, but also from a variety of other sources, including:
\bi
\item energy loss from cracks and un-instrumented regions of the detector,
\item energy loss from dead cells,
\item hot cells in the calorimeter that report an energy deposition
even if there isn't one,
\item mis-measurement in the electromagnetic
calorimeters, hadronic calorimeters or muon detectors and
\item the presence of mis-identified cosmic rays in events.
\ei
Thus, in order to have a solid grasp of expected $\eslt$ from
SM background processes, it will be necessary to have detailed knowledge of the
complete detector performance. As experience at the Tevatron suggests,
this complicated task may well take some time to complete. 
The same is likely to be true at the LHC, 
as many SM processes will have to be scrutinized first 
in order to properly calibrate the detector. 
For this reason, SUSY searches using $\eslt$
as a crucial requirement may well take rather longer than a 
year to provide reliable results.

On the other hand, if SUSY particles are relatively light, then 
production cross sections can be huge, and many new physics events may 
be generated in the first few  months of running. For instance,
for $m_{\tg}\sim 400$ GeV and heavy squarks, the expected gluino pair
cross sections are in the $10^5$ fb range. If 
$m_{\tg}\sim m_{\tq}\sim 400$ GeV, 
then production cross sections are even higher: of order $10^6$ fb! 
Thus, with just
0.1 fb$^{-1}$ of integrated luminosity, we might expect of order
$10^4-10^5$ new physics events to be recorded on tape if the gluino is in the
400 GeV range.

In this paper, we wish to examine if an early SUSY discovery might be made
{\it without} using $\eslt$ cuts. 
The key is to take advantage of the large production cross sections
of strongly interacting SUSY particles (the gluinos and squarks)
and their complex cascade decays.
Gluinos and squarks generally decay through a multi-step cascade of
decays\cite{cascade} 
until the LSP state is reached, so that SUSY signal events are 
expected to be rich in jet multiplicity, $b$-jet multiplicity,
isolated lepton multiplicity and sometimes large tau-jet multiplicity.
In addition, gauge mediated SUSY can lead to collider events 
with high isolated photon multiplicity. 
Thus, we would like to be able to use 
{\it detected objects} such as jets, $b$-jets and isolated leptons
to maximize signal over background, rather than {\it inferred} objects like
$\eslt$ which requires a complete detector knowledge.
Our main result in this paper is that we find a substantial 
reach for SUSY at the LHC by requiring multi-jet plus multi-lepton
events, without requiring the presence of $\eslt$.\footnote{
Similar signal calculations for models with a charged stable LSP have been performed in Ref. \cite{bis}.} 
By searching in this channel, one may be able to discover SUSY even before
the detectors are fully calibrated such that $\eslt$ is a useful variable
for background rejection. 

We use Isajet 7.76\cite{isajet} for the simulation of signal and 
background events at the LHC. A toy detector simulation is employed with
calorimeter cell size
$\Delta\eta\times\Delta\phi=0.05\times 0.05$ and $-5<\eta<5$. The HCAL
energy resolution is taken to be $80\%/\sqrt{E}+3\%$ for $|\eta|<2.6$ and
FCAL is $100\%/\sqrt{E}+5\%$ for $|\eta|>2.6$. 
The ECAL energy resolution
is assumed to be $3\%/\sqrt{E}+0.5\%$. We use a UA1-like jet finding algorithm
with jet cone size $R=0.4$ and require that $E_T(jet)>50$ GeV and
$|\eta (jet)|<3.0$. Leptons are considered
isolated if they have $p_T(e\ or\ \mu)>20$ GeV and $|\eta|<2.5$ with 
visible activity within a cone of $\Delta R<0.2$ of
$\Sigma E_T^{cells}<5$ GeV. The strict isolation criterion helps reduce
multi-lepton backgrounds from heavy quark ($c\bar c$ and $b\bar{b}$) production.

We identify a hadronic cluster with $E_T>50$ GeV and $|\eta(j)|<1.5$
as a $b$-jet if it contains a $B$ hadron with $p_T(B)>15$ GeV and
$|\eta (B)|<3$ within a cone of $\Delta R<0.5$ about the jet axis. We
adopt a $b$-jet tagging efficiency of 60\%, and assume that
light quark and gluon jets can be mis-tagged as $b$-jets with a
probability $1/150$ for $E_T\le 100$ GeV, $1/50$ for $E_T\ge 250$ GeV, 
with a linear interpolation for $100$ GeV$<E_T<$ 250 GeV\cite{xt}. 

For our initial analysis, we adopt the well-studied \sps1ap benchmark
point\cite{snops}, which occurs in the minimal supergravity (mSUGRA) model
with parameters $m_0=70$ GeV, $m_{1/2}=250$ GeV, $A_0=-300$ GeV, 
$\tan\beta =10$, $\mu >0$ and $m_t=171$ GeV.
Here, $m_0$ is a common GUT scale scalar soft breaking mass, $m_{1/2}$
is a common GUT scale gaugino mass, $A_0$ is a common GUT scale trilinear
soft term and $\tan\beta$ is the ratio of Higgs vevs. The parameter
$\mu$ occurs in the superpotential; its magnitude, but not its sign, is determined
by requiring a radiative breakdown of electroweak symmetry.
The sparticle mass spectrum is generated by the Isajet 7.76 program, 
which adopts an iterative approach to solving the MSSM RGEs 
using two-loop RGEs and complete 1-loop sparticle mass radiative corrections.
The \sps1ap point leads to a spectrum with $m_{\tg} =608$ GeV, while squark masses
tend to be in the 550 GeV range. The gluinos and squarks then cascade decay via
a multitude of modes leading to events with high jet, $b$-jet, isolated lepton
and tau lepton multiplicity.

In addition, we have generated background events using Isajet for
QCD jet production (jet-types include $g$, $u$, $d$, $s$, $c$ and $b$
quarks) over five $p_T$ ranges as shown in Table \ref{tab:bg}. 
Additional jets are generated via parton showering from the initial and final state
hard scattering subprocesses.
We have also generated backgrounds in the $W+jets$, $Z+jets$, 
$t\bar{t}(171)$ and $WW,\ WZ,\ ZZ$ channels at the rates shown in 
Table \ref{tab:bg}. The $W+jets$ and $Z+jets$ backgrounds
use exact matrix elements for one parton emission, but rely on the 
parton shower for subsequent emissions.
\begin{table}
\begin{center}
\begin{tabular}{lccc}
\hline
process & events & $\sigma$ (fb) & cuts \c1p$+\ge 3\ell$ (fb)  \\
\hline
QCD ($p_T:50-100$ GeV) & $10^6$ & $2.6\times 10^{10}$ & --\\
QCD ($p_T:100-200$ GeV) & $10^6$ & $1.5\times 10^{9}$ & -- \\
QCD ($p_T:200-400$ GeV) & $10^6$ & $7.3\times 10^{7}$ & -- \\
QCD ($p_T:400-1000$ GeV) & $10^6$ & $2.7\times 10^{6}$ & -- \\
QCD ($p_T:1000-2400$ GeV) & $10^6$ & $1.5\times 10^{4}$ & -- \\
$W+jets; W\to e,\mu,\tau$ $(p_T(W):100-4000$ GeV)& $5\times 10^5$ & 
$3.9\times 10^{5}$ & 0.8 \\
$Z+jets; Z\to \tau\bar{\tau},\ \nu s$ $(p_T(Z):100-3000$ GeV) & $5\times 10^5$ & 
$1.4\times 10^{5}$ & 0.3 \\
$t\bar{t}$ & $3\times 10^6$ & $5.1\times 10^{5}$ & 5.1 \\
$WW,ZZ,WZ$ & $5\times 10^5$ & $8.0\times 10^{4}$ & --  \\
signal (\sps1ap: $m_{\tg}=608$ GeV) & $2.5\times 10^5$ & $4.7\times 10^4$ 
& $46.6$ \\
\hline
\end{tabular}
\caption{Events generated and cross sections for various SM background 
processes plus the \sps1ap case study.
The \c1p cuts are specified in Eqns. ($1 - 3$).}
\label{tab:bg}
\end{center}
\end{table}

We begin by applying a set of pre-cuts to our event samples. These cuts,
known as set C1 in Ref. \cite{gabe}, were used for studying gluino mass 
determination in the focus point region of mSUGRA. 
Here, we abandon the $\eslt>(100\ {\rm GeV},0.2M_{eff})$ cut
and call the new set of cuts \c1p.
\\
\\
\textbf{C1$^\prime$ cuts:}
\bea
n(jets) &\ge & 4,\\
E_T(j1,j2,j3,j4)& \ge & 100,\ 50,50,50\ {\rm GeV},\\
S_T &\ge &0.2 .
\label{c1pcuts}
\eea
We will also make use of the {\it augmented} effective mass
$A_T=\eslt +\sum_{jets} E_T(j)+\sum_{leptons} E_T(\ell)$.
Here, $\ell$ stands for either $e$ or $\mu$.
If we remove $\eslt$ from $A_T$, we will call the new variable \atp.
$S_T$ is transverse sphericity\footnote{Sphericity is defined, {\it e.g.} in
{\it Collider Physics}, V. Barger and R. J. N. Phillips (Addison Wesley, 1987)
Here, we restrict its construction to using only transverse quantities, 
as is appropriate for a hadron collider.}.
The event rates in fb are listed before cuts in column 3 of Table \ref{tab:bg}.

\begin{figure}[htbp]
\begin{center}
\vskip .2cm
\includegraphics[width=0.69\textwidth]{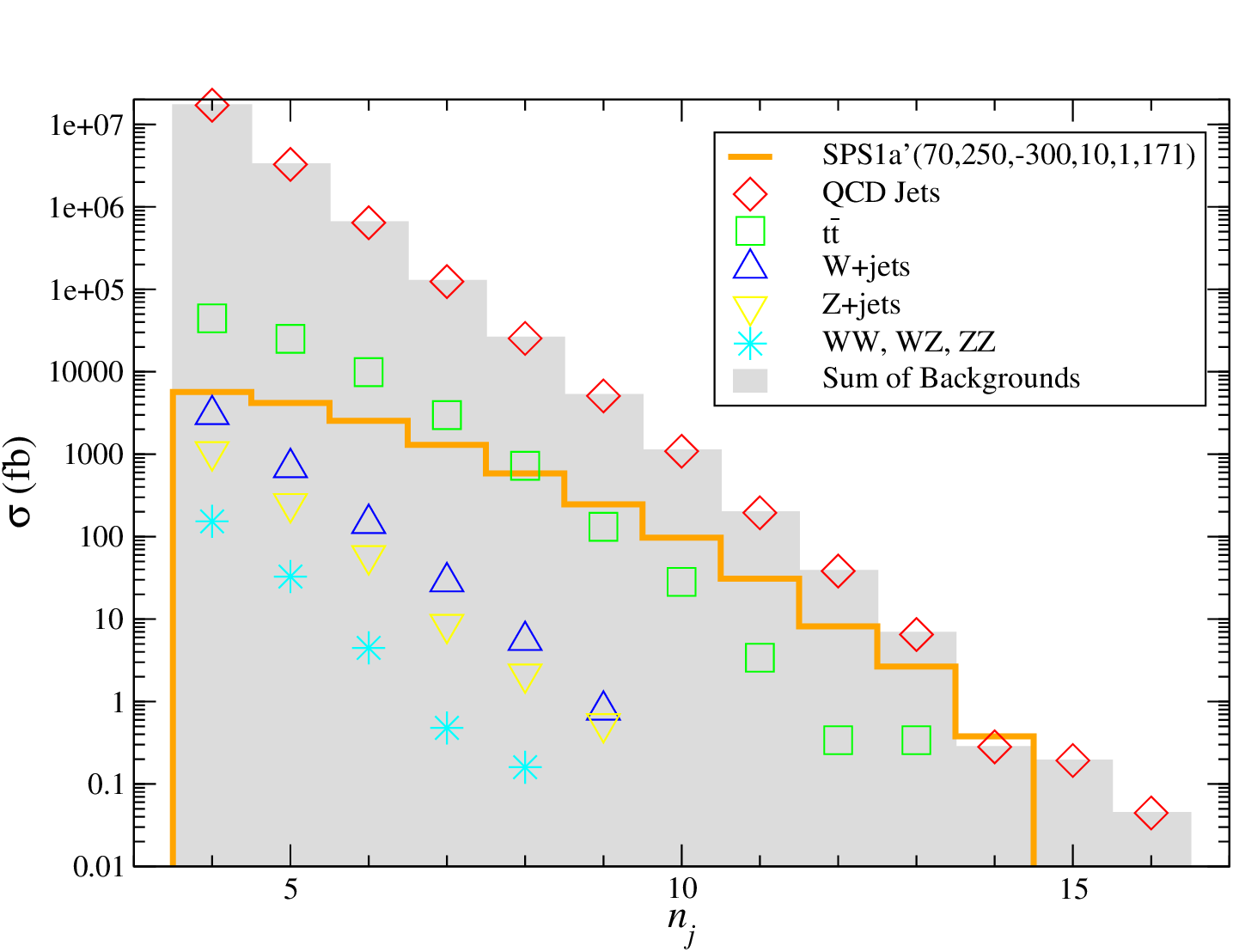}
\caption{Plot of jet multiplicity from SUSY collider events from \sps1ap
after cuts \c1p . We also plot the histograms of various SM backgrounds, 
plus the total SM background (gray histogram).
}
\label{fig:nj}
\end{center}
\end{figure}

In Fig. \ref{fig:nj}, we plot the resulting jet multiplicity $n_j$
after cuts \c1p for the \sps1ap benchmark (orange histogram) along with 
the various SM backgrounds. 
The gray histogram gives the sum of all SM backgrounds. We see immediately
that SM background, dominated by QCD multi-jet production, dominates
out to very high jet multiplicity.

\begin{figure}[htbp]
\begin{center}
\includegraphics[width=0.69\textwidth]{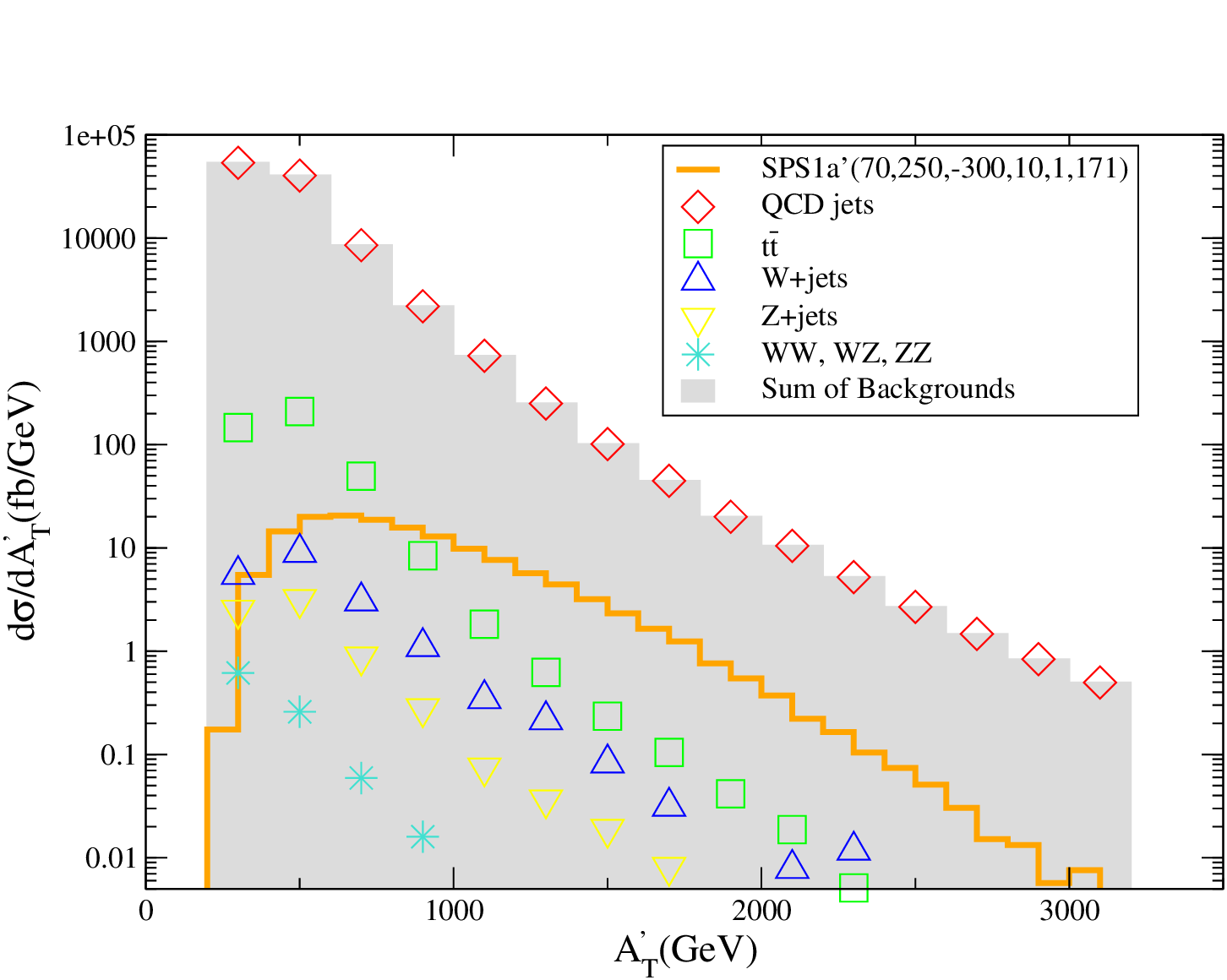}
\caption{Plot of augmented effective mass \atp (without $\eslt$) 
from SUSY collider events from \sps1ap
after cuts \c1p . We also plot the histograms of various SM backgrounds, 
plus the total SM background (gray histogram).
}
\label{fig:atp}
\end{center}
\end{figure}

In Fig. \ref{fig:atp}, we plot the augmented effective mass (minus
the $\eslt$ component) \atp . When the $\eslt$ cut is used in cut
set C1, then signal generally emerges from background at some large
value of $A_T$ which is somewhat correlated with the values of
$m_{\tg}$ and $m_{\tq}$\cite{frank}. In this case, with no $\eslt$ cut,
the signal is hopelessly below the summed BG distribution.

\begin{figure}[htbp]
\begin{center}
\includegraphics[width=0.69\textwidth]{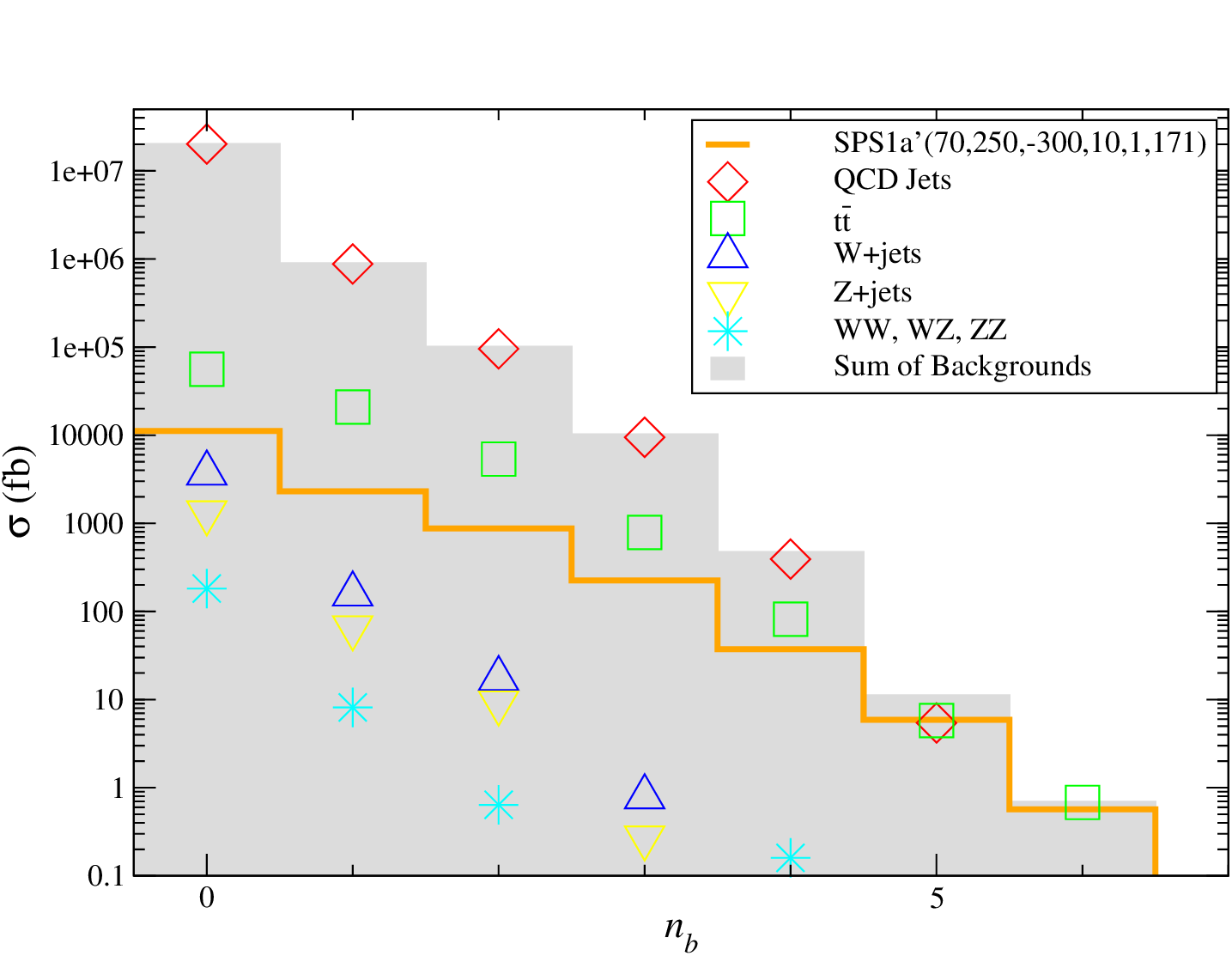}
\caption{Plot of $b$-jet multiplicity $n_b$ from LHC SUSY events from 
\sps1ap after cuts \c1p . 
We also plot the histograms of various SM backgrounds, 
plus the total SM background (gray histogram).
}
\label{fig:nbj}
\end{center}
\end{figure}

In Fig. \ref{fig:nbj}, we plot the multiplicity of tagged $b$-jets $n_b$ in 
events after cuts \c1p . We see that out to $n_b=5$, SM background from
QCD jet production-- including both $b\bar{b}$ production, 
parton shower production from $g\to b\bar{b}$  and also jets faking
a $b$-jet-- dominates the signal.  

\begin{figure}[htbp]
\begin{center}
\includegraphics[width=0.69\textwidth]{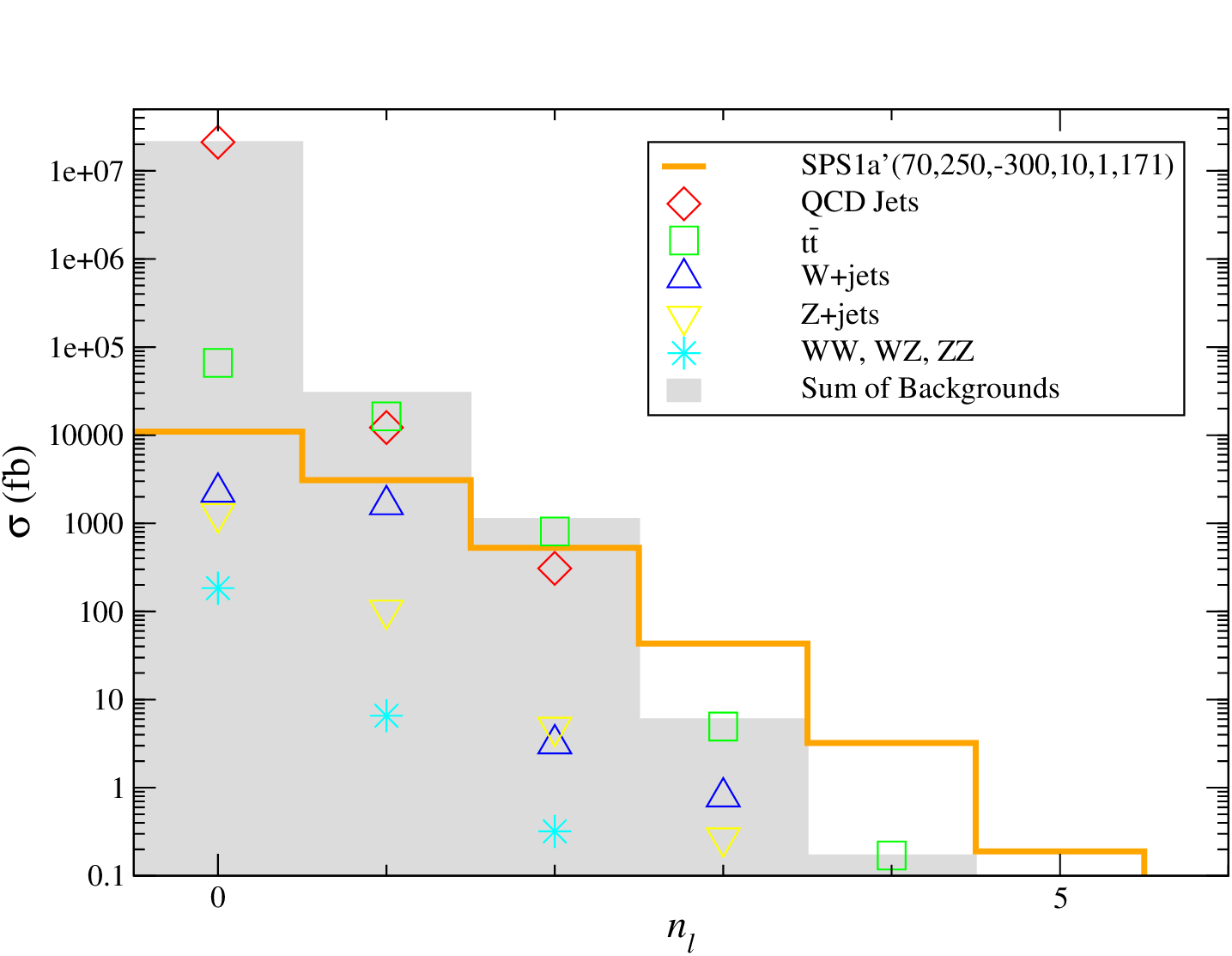}
\caption{Plot of isolated lepton multiplicity $n_\ell$ from LHC  
SUSY events from \sps1ap after cuts \c1p . 
We also plot the histograms of various SM backgrounds, 
plus the total SM background (gray histogram).
}
\label{fig:nl}
\end{center}
\end{figure}

In Fig. \ref{fig:nl}, we plot the multiplicity of isolated leptons $n_\ell$
for benchmark point \sps1ap and SM background. Here we see that
at low values of  $n_\ell =0$ or 1, signal is dominated by BG. However,
at $n_\ell =2$, signal is above QCD BG, and only below $t\bar{t}$ BG.
By the time we require $n_\ell =3$, SM background is well below signal.
In this case, it is clear that we can gain good BG rejection by
requiring the cut set \c1p , plus $n_\ell \ge 3$. The remaining signal is
at the 40-50 fb level, which should be adequate for discovery if
0.1-1 fb$^{-1}$ of integrated luminosity is obtained. 
The dominant
background comes from $t\bar{t}$ production. An early verification of 
$t\bar{t}$ production via its one and two lepton signatures should allow 
for a solid calibration of this most important background.

To gain an estimate of the LHC reach using cuts \c1p plus $\ge 3\ell$, 
we set $m_0 =200$ GeV (lighter squarks) and $m_0=1000$ GeV (heavy squarks)
and vary $m_{1/2}$ from 170 to 500 GeV. We also take $A_0=0$, $\tan\beta =10$
and $\mu >0$. We plot the resulting signal cross section as a function of 
$m_{\tg}$ rather than $m_{1/2}$ in order to explicitly show the reach
in terms of a measureable parameter. 
We also plot the model line of $m_0=70$ GeV and $A_0=-300$ GeV (with $\tan\beta =10$ and 
$\mu >0$) which contains the \sps1ap point.
The $5\sigma$ BG level is shown
for 0.1 and 1 fb$^{-1}$ of integrated luminosity. 
(The $5\sigma$ level for 0.1 fb$^{-1}$ is $\sim 40$ fb, so would only correspond to four events.)
While the total signal
cross section for the red curve (with $m_0=200$ GeV) is larger than that for 
the blue curve (for $m_0=1000$ GeV), the cross section after cuts is
actually larger for the large $m_0$ case at low $m_{1/2}$. 
This is because
in this region, around $m_0=m_{1/2}\sim 200$ GeV, the $\tz_2$ branching fraction 
to leptons $\tz_2\to\tz_1\ell\bar{\ell}$ is suppressed due to destructive 
interference in the $Z$ and slepton mediated decay processes.
In the high $m_0$ case, 3-body decay of $\tz_2$ via the $Z^*$ is always dominant. 
However, in all cases, we see the $5\sigma$ reach extends to $m_{\tg}\sim 700-750$
GeV for 0.1 fb$^{-1}$ of integrated luminosity, and out to $m_{\tg}\sim 1$ TeV
for 1 fb$^{-1}$ of integrated luminosity.
This would represent a significant leap in experimental sensitivity to $m_{\tg}$
which could be obtained at relatively low LHC integrated luminosity, 
while not using $\eslt$ cuts. 
\begin{figure}[htbp]
\begin{center}
\vskip .2cm
\includegraphics[width=0.69\textwidth]{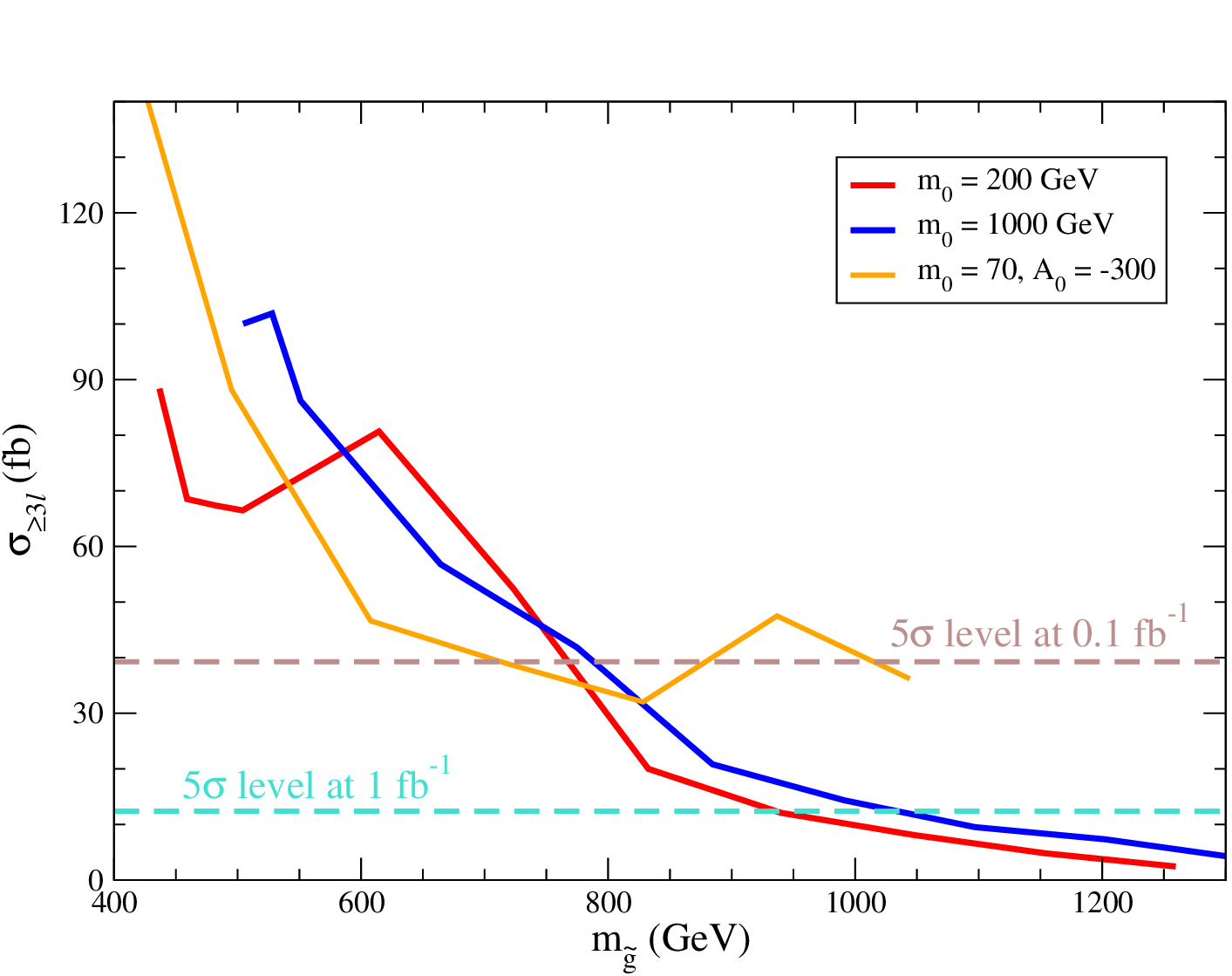}
\caption{Plot of signal cross section from mSUGRA model versus
$m_{\tg}$ after cuts \c1p and $n_\ell\ge 3$, for $m_0=200$ and
1000 GeV. We also take $A_0=0$, $\tan\beta =10$, $\mu >0$ and
$m_t =171$ GeV. We also plot the $5\sigma$ background level 
for 0.1 and 1 fb$^{-1}$ of integrated luminosity.
}
\label{fig:mgl}
\end{center}
\end{figure}

\begin{figure}[htbp]
\begin{center}
\includegraphics[width=0.69\textwidth]{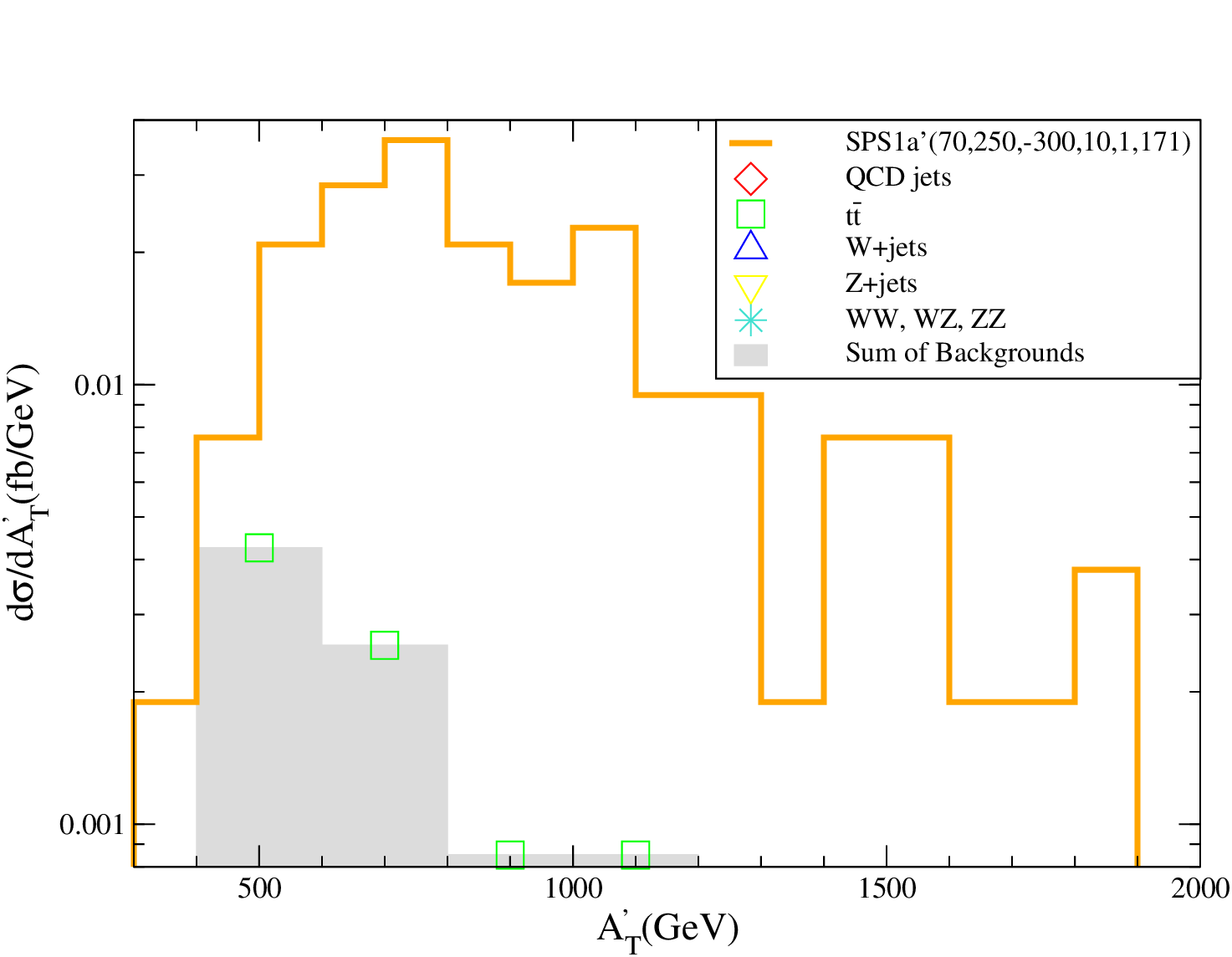}
\caption{Distribution in variable \atp 
from SUSY events from \sps1ap after cuts \c1p plus $\ge 3\ell$ plus
$\ge 1$ $b$-jet. 
We also plot the remaining SM backgrounds (gray histogram).
}
\label{fig:bat}
\end{center}
\end{figure}

One possible criticism of our results so far is that we use only leading order
cross sections as calculated by Isajet. However, we expect that NLO
total cross sections for both signal and background to be somewhat enhanced
beyond the LO Isajet results, so we would expect our overall conclusion to remain
valid qualitatively. Indeed, it is expected that the major SM processes will
be measured to high accuracy at LHC already at low luminosity, so that a good
background calibration should be at hand. 
A second criticism could be that there are additional background processes to be checked.
These would include $2\to n$ processes such as $\ttbar\ttbar$, $\ttbar V$, 
$\ttbar VV$, $VVV$ and $VVVV$ production, where $V=W^\pm$ or $Z$. 
While these processes occur at higher 
order in perturbation theory, they do offer the possibility to generate multi-lepton
final states rather efficiently. We hope to address these in a future work.
A third criticism might be that
we have not taken any ``jet faking a lepton'' probability into account. 
This possibility is detector and lepton flavor dependent. 
However, if it turns out to be a problem, we can only note that our cuts so far
have been rather minimal, and can easily be extended. For instance, 
requiring the presence of one $b$-jet in each event will severely reduce
$W+$jets and $Z$+jets BG. Then, plotting the distribution in \atp
will allow signal to emerge from $\ttbar$ and other backgrounds.
This is illustrated in Fig. \ref{fig:bat}. 

We also note here that if a SUSY signal is found in the $\ge 4$ jets plus
$\ge 3\ell$ sample, then the resulting event sample may be used for
precision sparticle mass measurements just as in the case where one requires
jets $+\eslt$. As an example, we examine all events passing cuts \c1p and
$\ge 3\ell$ for benchmark \sps1ap and plot the invariant mass of all
opposite sign/same flavor (OS/SF) dilepton pairs. In this case, we 
expect a mass edge\cite{mll} at 
$m(\ell\bar{\ell})=m_{\tz_2}\sqrt{1-\frac{m_{\tell}^2}{m_{\tz_2}^2}}
\sqrt{1-\frac{m_{\tz_1}^2}{m_{\tell}^2}}=82.3$ GeV 
(since here $m_{\tz_2}=183.0$, $m_{\tell_R}=123.3$ GeV and $m_{\tz_1}=97.8$ GeV). 
The mass edge is evident from the plot, and serves as a starting point for
further sparticle mass reconstruction. 
\begin{figure}[htbp]
\begin{center}
\vskip .4cm
\includegraphics[width=0.69\textwidth]{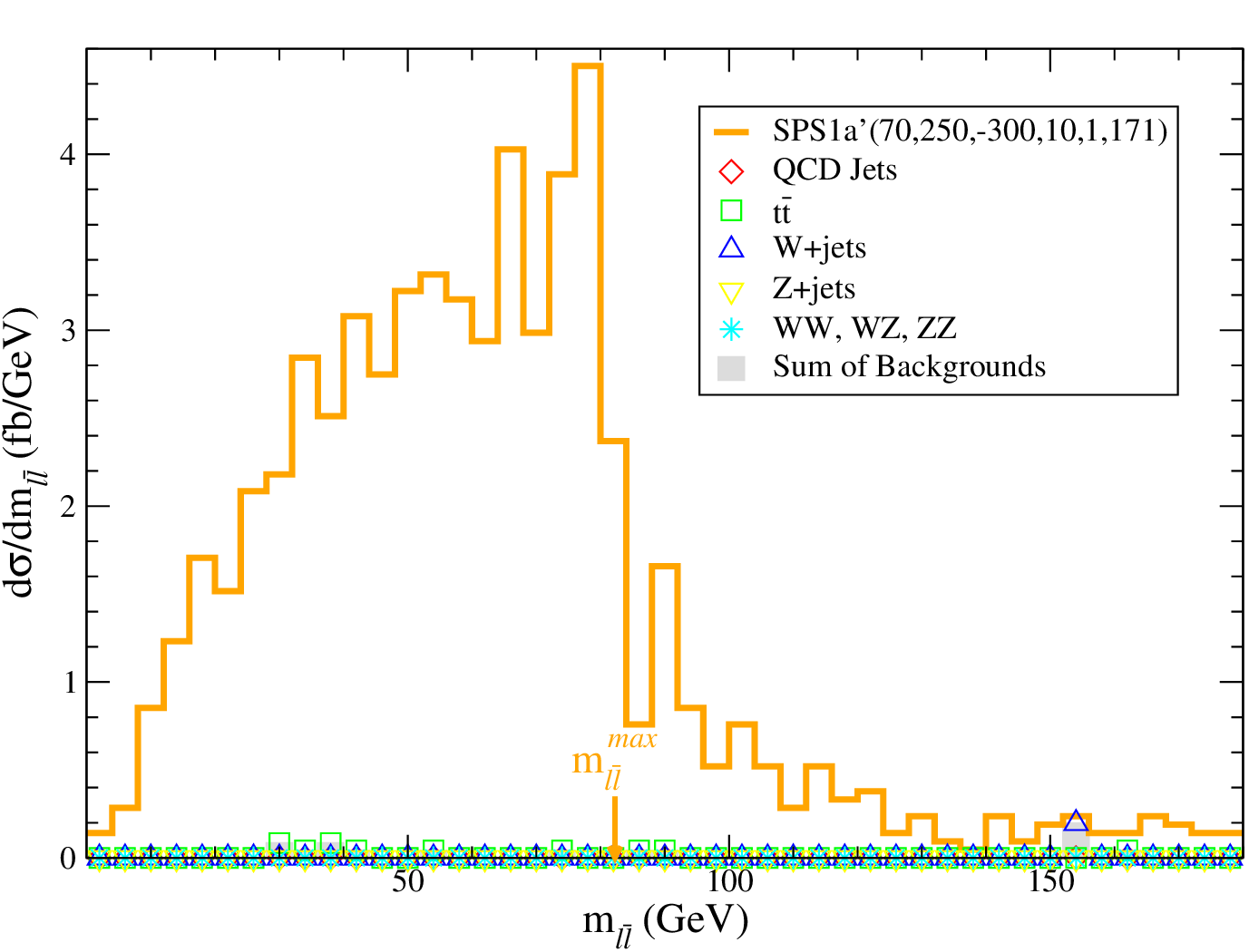}
\caption{Plot of OS/SF dilepton invariant mass from
SUSY events from \sps1ap after cuts \c1p plus $\ge 3\ell$. 
We also plot the remaining SM backgrounds (gray histogram).
}
\label{fig:mll}
\end{center}
\end{figure}

\begin{figure}[htbp]
\begin{center}
\vskip .2cm
\includegraphics[width=0.67\textwidth]{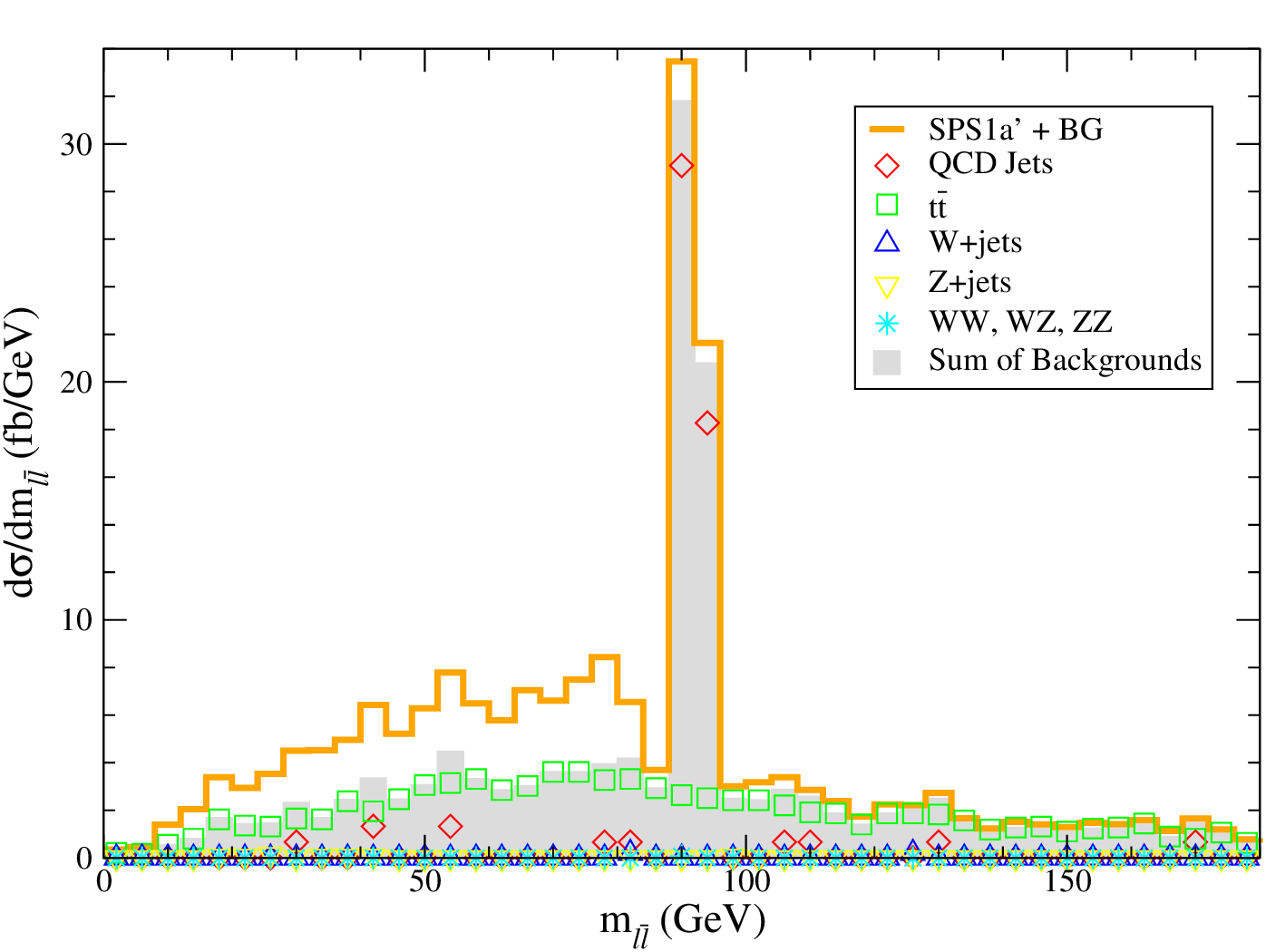}
\vskip 1cm
\includegraphics[width=0.67\textwidth]{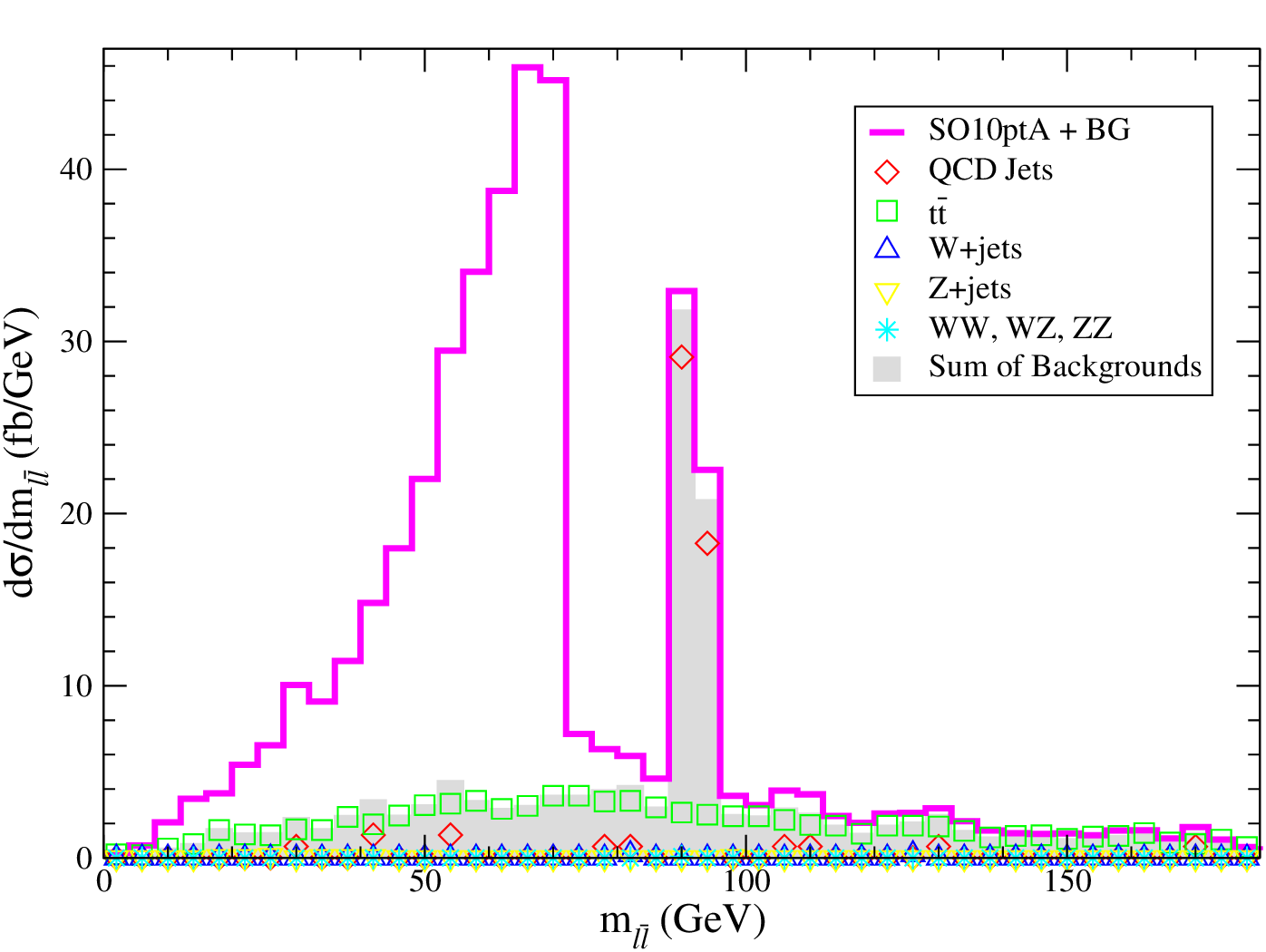}
\caption{Plot of OS/SF dilepton invariant mass from
SUSY events from {\it a}) \sps1ap 
and {\it b}) point A of $SO(10)$ benchmarks after cuts \c1p plus a
OS/SF pair of leptons.
We also plot the remaining SM backgrounds (gray histogram).
}
\label{fig:llb}
\end{center}
\end{figure}

While the requirement of jets $+\ge 3\ell$ works to see $m_{\tg}\alt 750$ GeV
with just 0.1 fb$^{-1}$ of integrated luminosity, it is possible to see SUSY signals 
with even lower lepton multiplicities. To illustrate, we examine SUSY models 
which give rise to the distinctive dilepton invariant mass edge from $\tz_2$ decay 
to $\ell\bar{\ell}\tz_1$. In this case, we require cuts \c1p plus an OS/SF
lepton pair. We plot out in Fig. \ref{fig:llb}{\it a}) the resultant distribution in
$m(\ell\bar{\ell})$ for both \sps1ap and SM BG.
We see a continuum of background, along with a $Z$ peak. The BG $Z$ peak arises
because Isajet includes $W$ and $Z$ radiation in its parton shower algorithm.
The orange histogram shows the sum of signal plus BG, and the OS/SF dilepton mass
edge clearly stands out below the $Z$ peak.
A further example comes from $SO(10)$ benchmark point A suggested in Ref. \cite{so10},
which includes a $\sim 400$ GeV gluino. In this case, the large $\tg\tg$
cross section allows signal to stand out even more abruptly from  SM background.

{\it Conclusions:} In the very early run of the LHC $pp$ collider, 
it may not be possible to use $\eslt$ as a discrimination variable due to
detector calibration issues. We show here that a substantial reach for
gluino and squark production followed by cascade decays can be gained
by requiring events with large jet and isolated lepton multiplicity, but with
no requirement on $\eslt$. In the mSUGRA model with a low and high value of
$m_0$, an LHC reach for $m_{\tg}$ of 750 (1000) GeV is found with 0.1 (1) fb$^{-1}$
of integrated luminosity by requiring $\ge $4 jets plus $\ge 3$ isolated leptons.
If enough signal events are found, then some kinematic reconstruction of 
sparticle masses should be possible as in the cases where large $\eslt$ is
required. SUSY signal can also be seen above SM BG if just two OS/SF leptons
are required, especially in the case where there is a distinctive 
kinematic dilepton invariant mass edge.

\section*{Acknowledgments}

This work was supported in part by the U.S.~Department of Energy under 
grant No. DE-FG02-95ER40896. 
We thank Y. Gershtein and X. Tata for comments on the manuscript.

%

%
\end{document}